\documentclass[prd,showpacs,showkeys,floatfix,twocolumn,amsmath,amssymb,floatfix]{revtex4}
\usepackage{graphicx,color,dcolumn,booktabs,bm}
\usepackage{subfigure}
\usepackage{longtable,lscape}
\usepackage{amssymb}
\usepackage{indentfirst}
\usepackage{epsfig}
\usepackage{feynmf}   
\usepackage{epstopdf}   
\usepackage{slashed}  
\usepackage{cases}
\usepackage{color}
\usepackage{multirow}
\usepackage{graphicx,color,dcolumn,booktabs,bm}
\usepackage[colorlinks, citecolor=blue,anchorcolor=red,menucolor=red, linkcolor=red,filecolor=red,runcolor=red,urlcolor=blue,frenchlinks=red]{hyperref}
\begin{document}
\title{Possible interpretations of the $P_c(4312)$, $P_c(4440)$, and $P_c(4457)$}
%

\author{Hua-Xing Chen$^1$}
\email{hxchen@buaa.edu.cn}
\author{Wei Chen$^2$}
\email{chenwei29@mail.sysu.edu.cn}
\author{Shi-Lin Zhu$^{3,4,5}$}
\email{zhusl@pku.edu.cn}
\affiliation{
$^1$School of Physics, Beihang University, Beijing 100191, China \\
$^2$School of Physics, Sun Yat-Sen University, Guangzhou 510275, China \\
$^3$School of Physics and State Key Laboratory of Nuclear Physics and Technology, Peking University, Beijing 100871, China \\
$^4$Collaborative Innovation Center of Quantum Matter, Beijing 100871, China \\
$^5$Center of High Energy Physics, Peking University, Beijing 100871, China
}
\begin{abstract}
Based on our previous QCD sum rule studies on hidden-charm pentaquark states, we discuss possible interpretations of the $P_c(4312)$, $P_c(4440)$, and $P_c(4457)$, which were recently observed by LHCb. Our results suggest that the $P_c(4312)$ can be well interpreted as the $[\Sigma_c^{++} \bar D^-]$ bound state with $J^P = 1/2^-$, while the $P_c(4440)$ and $P_c(4457)$ can be interpreted as  the $[\Sigma_c^{+} \bar D^0]$ bound state with $J^P = 1/2^-$, the $[\Sigma_c^{*++} \bar D^{-}]$ and $[\Sigma_c^{+} \bar D^{*0}]$ bound states with $J^P = 3/2^-$, or the $[\Sigma_c^{*+} \bar D^{*0}]$ bound state with $J^P = 5/2^-$. We propose to measure their spin-parity quantum numbers to verify these assignments.
\end{abstract}
\maketitle

\pagenumbering{arabic}

$\\$
{\it Introduction.}
Very recently, the LHCb Collaboration discovered a new enhancement, $P_c(4312)$, in the $J/\psi p$ invariant mass spectrum of the $\Lambda_b\to J/\psi p K$ decays~\cite{lhcb}. At the same time, they separated the $P_c(4450)$ into two structures, $P_c(4440)$ and $P_c(4457)$. This experiment~\cite{lhcb} was based on their previous one performed in 2015~\cite{Aaij:2015tga}, where the famous hidden-charm pentaquark candidates, $P_c(4380)$ and $P_c(4450)$, were first observed. All the above structures contain at least five quarks $uud c \bar c$, so they are perfect candidates of pentaquark states. Together with many charmonium-like $XYZ$ states~\cite{pdg}, their studies have improved our understanding of the non-perturbative behaviors of the strong interaction at the low energy region. However, there is still a long way to fully understand how the strong interaction binds quarks, gluons, and hadrons together, and exotic hadrons will continuously be one of the most intriguing research topics in hadron physics.

In the new LHCb experiment~\cite{lhcb} the following resonance parameters were measured:
\begin{eqnarray}
\nonumber     P_c^+(4312):     M &=& 4311.9 \pm  0.7 ^{+6.8}_{-0.6}  \mbox{ MeV} \, ,
\\ \nonumber            \Gamma &=&    9.8 \pm  2.7 ^{+3.7}_{-4.5}  \mbox{ MeV} \, ,
\\ \nonumber  P_c^+(4440):     M &=& 4440.3 \pm  1.3 ^{+4.1}_{-4.7}  \mbox{ MeV} \, ,
\\ \nonumber            \Gamma &=&   20.6 \pm  4.9 ^{+8.7}_{-10.1} \mbox{ MeV} \, ,
\\ \nonumber  P_c^+(4457):     M &=& 4457.3 \pm  0.6 ^{+4.1}_{-1.7}  \mbox{ MeV} \, ,
\\                      \Gamma &=&    6.4 \pm  2.0 ^{+5.7}_{-1.9}  \mbox{ MeV} \, .
\label{experiment}
\end{eqnarray}
Hence, these three structures are quite narrow, and can be clearly seen in the $J/\psi p$ invariant mass spectrum.
As discussed by LHCb~\cite{lhcb}, the $P_c(4312)$, $P_c(4440)$, and $P_c(4457)$ are just below the $\Sigma_c^+ \bar D^0$ and $\Sigma_c^+ \bar D^{*0}$ thresholds (4318 MeV and 4460 MeV~\cite{pdg}, respectively), so they can be naturally interpreted as the bound states composed of charmed baryons and anti-charmed mesons, whose existence had been predicted in several theoretical studies~\cite{Wu:2010jy,Wang:2011rga,Yang:2011wz,Wu:2012md,Li:2014gra,Karliner:2015ina} before the LHCb-2015 experiment~\cite{Aaij:2015tga}; while after this experiment~\cite{Aaij:2015tga}, lots of theoretical studies were performed to explain the nature of the $P_c(4380)$ and $P_c(4450)$, such as meson-baryon molecules~\cite{Chen:2015loa,Chen:2015moa,Roca:2015dva,He:2015cea,Meissner:2015mza,Azizi:2016dhy,Guo:2019kdc}, compact diquark-diquark-antiquark or diquark-triquark pentaquarks~\cite{Maiani:2015vwa,Lebed:2015tna,Wang:2015epa}, and kinematical effects related to the triangle singularity~\cite{Guo:2015umn,Liu:2015fea,Bayar:2016ftu}, etc.

We have applied the method of QCD sum rules~\cite{Shifman:1978bx,Reinders:1984sr} to systematically study the hidden-charm pentaquarks in Refs.~\cite{Chen:2015moa,Chen:2016otp}. Based on the new experimental information~\cite{lhcb} as well as our previous theoretical studies~\cite{Chen:2015moa,Chen:2016otp}, we shall discuss several possible interpretations of the $P_c(4312)$, $P_c(4440)$, and $P_c(4457)$ in this letter.

$\\$
{\it The first possible interpretation.}
In Ref.~\cite{Chen:2015moa} we applied the method of QCD sum rules and studied the $P_c(4380)$ and $P_c(4450)$ as exotic hidden-charm pentaquarks composed of charmed baryons and anti-charmed mesons. 
This study was later expanded in Ref.~\cite{Chen:2016otp}, where we systematically constructed all the possible local hidden-charm pentaquark currents with spin $J = {1\over2}/{3\over2}/{5\over2}$ and quark contents $uud c \bar c$, and investigated them using the method of QCD sum rules.

Especially, in the abstract of Ref.~\cite{Chen:2016otp} we wrote that:{\it``...we also find a) the lowest-lying hidden-charm pentaquark state of $J^P = 1/2^-$ has the mass $4.33^{+0.17}_{-0.13}$ GeV, while the one of $J^P = 1/2^+$ is significantly higher, that is around $4.7-4.9$ GeV; b) the lowest-lying hidden-charm pentaquark state of $J^P = 3/2^-$ has the mass $4.37^{+0.18}_{-0.13}$ GeV, consistent with the $P_c(4380)$ of $J^P = 3/2^-$, while the one of $J^P = 3/2^+$ is also significantly higher, that is above $4.6$ GeV; c) the hidden-charm pentaquark state of $J^P = 5/2^-$ has a mass around $4.5-4.6$ GeV, slightly larger than the $P_c(4450)$ of $J^P = 5/2^+$.''}
Comparing these values with Eqs.~(\ref{experiment}), we arrive at the first possible interpretation that {\it (A) the $P_c(4312)$ is the hidden-charm pentaquark state with $J^P = 1/2^-$, while the $P_c(4440)$ and $P_c(4457)$ may be the two with $J^P = 3/2^-$ or/and $5/2^-$}.

\renewcommand{\arraystretch}{1.5}
\begin{table*}[]
\begin{center}
\caption{Mass predictions for the hidden-charm pentaquark states with spin $J = {1\over2}/{3\over2}/{5\over2}$ and quark contents $uud c \bar c$, taken from Ref.~\cite{Chen:2016otp}. We summarize here all the mass predictions that are extracted from single currents and less than 4.5~GeV.}
\begin{tabular}{ccc|cc|cc}
\toprule[1pt]\toprule[1pt]
~~\mbox{Current}~~ & ~~\mbox{Defined in}~~ & ~~\mbox{Structure}~~ & \mbox{$s_0$ [GeV$^2$]} & \mbox{Borel Mass [GeV$^2$]} & ~~\mbox{Mass [GeV]}~~ & ~~\mbox{($J$, $P$)}~~
\\ \midrule[1pt]
$\xi_{14}$        & Eq.~(\ref{def:xi14})  & $[\Sigma_c^{+} \bar D^0]$      & $20 - 24$ & $4.12 - 4.52$ & $4.45^{+0.17}_{-0.13}$ & ($1/2,-$)
\\
$\psi_2$          & Eq.~(\ref{def:psi2})  & $[\Sigma_c^{++} \bar D^-]$    & $19 - 23$ & $3.95 - 4.47$ & $4.33^{+0.17}_{-0.13}$ & ($1/2,-$)
\\ \midrule[1pt]
$\xi_{33\mu}$               & Eq.~(\ref{def:xi33mu})  & $[\Sigma_c^+ \bar D^{*0}]$    & $20 - 24$ & $3.97 - 4.41$ & $4.46^{+0.18}_{-0.13}$ & ($3/2,-$)
\\
$\psi_{2\mu}$               & Eq.~(\ref{def:psi2mu})  & $[\Sigma_c^{*++} \bar D^-]$    & $20 - 24$ & $3.88 - 4.41$ & $4.45^{+0.16}_{-0.13}$ & ($3/2,-$)
\\
$\psi_{9\mu}$               & Eq.~(\ref{def:psi9mu})  & $[\Sigma_c^{++} \bar D^{*-}]$    & $19 - 23$ & $3.94 - 4.27$ & $4.37^{+0.18}_{-0.13}$ & ($3/2,-$)
\\ \midrule[1pt]
$\xi_{13\mu\nu}$     & Eq.~(\ref{def:xi13munu})  & $[\Sigma_c^{*+} \bar D^{*0}]$  & $20 - 24$ & $3.51 - 4.00$ & $4.50^{+0.18}_{-0.12}$ & ($5/2,-$)
\\ \bottomrule[1pt]\bottomrule[1pt]
\end{tabular}
\label{tab:result}
\end{center}
\end{table*}

However, this picture is quite rough and can not naturally explain the small mass difference between the $P_c(4440)$ and $P_c(4457)$, which is just about 17 MeV. To understand this mass splitting, we turn to carefully examine their internal structures. Actually, this can be well investigated and described by using hadronic interpolating currents within the method of QCD sum rules.

$\\$
{\it The second and third possible interpretations.} In Ref.~\cite{Chen:2016otp} we found that the internal structure of hidden-charm pentaquark states is quite complicated. We constructed hundreds of interpolating currents to reflect this, from which we derived some mass predictions. We collect all the mass predictions that are extracted from single currents and less than 4.5~GeV, and summarize them in Table~\ref{tab:result}. They are extracted using the following interpolating currents
\begin{eqnarray}
\xi_{14} &=& [\epsilon^{abc} (u^T_a C \gamma_\mu d_b) \gamma_\mu \gamma_5 c_c] [\bar c_d \gamma_5 u_d] \, ,
\label{def:xi14}
\\ \psi_2 &=& [\epsilon^{abc} (u^T_a C \gamma_\mu u_b) \gamma_\mu \gamma_5 c_c] [\bar c_d \gamma_5 d_d] \, ,
\label{def:psi2}
\\ \xi_{33\mu} &=& [\epsilon^{abc} (u^T_a C \gamma_\nu d_b) \gamma_\nu \gamma_5 c_c] [\bar c_d \gamma_\mu u_d] \, ,
\label{def:xi33mu}
\\ \psi_{2\mu} &=& [\epsilon^{abc} (u^T_a C \gamma_\mu u_b) c_c] [\bar c_d \gamma_5 d_d] \, ,
\label{def:psi2mu}
\\ \psi_{9\mu} &=& [\epsilon^{abc} (u^T_a C \gamma_\nu u_b) \gamma_\nu \gamma_5 c_c] [\bar c_d \gamma_\mu d_d] \, ,
\label{def:psi9mu}
\\ \xi_{13\mu\nu} &=& [\epsilon^{abc} (u^T_a C \gamma_\mu d_b) c_c] [\bar c_d \gamma_\nu u_d] + \{ \mu \leftrightarrow \nu \} \, ,
\label{def:xi13munu}
\end{eqnarray}
where $u,d,c$ represent the {\it up}, {\it down}, and {\it charm} quarks, respectively, and the subscripts $a,b,c,d$ are color indices.
The above currents have the negative parity, but their mirror currents with the positive parity (such as $\gamma_5 \xi_{14}$, etc.) lead to the same QCD sum rule results. This is because each of them can couple to both the positive- and negative-parity pentaquark states, and we need to determine the parity of the state through the derived sum rule equations. See Refs.~\cite{Chung:1981cc,Jido:1996ia,Ohtani:2012ps,Chen:2015moa,Chen:2016otp} for detailed analyses.

From Table~\ref{tab:result}, we find four mass predictions, 4.33~GeV~($1/2^-$), 4.45~GeV~($1/2^-$), 4.45~GeV~($3/2^-$), and 4.46~GeV~($3/2^-$), which are almost the same as those listed in Eqs.~(\ref{experiment}). Accordingly, we arrive at the second and third possible interpretations that
{\it (B) the $P_c(4312)$ and $P_c(4440)$ are the two hidden-charm pentaquark states with $J^P = 1/2^-$, while the $P_c(4457)$ is the one with $J^P = 3/2^-$};
{\it (C) the $P_c(4312)$ is the hidden-charm pentaquark state with $J^P = 1/2^-$, while the $P_c(4440)$ and $P_c(4457)$ are the two with $J^P = 3/2^-$}.

It is usually not easy to understand the QCD sum rule results for multiquark states, because we still do not well understand the relations between interpolating currents and their relevant hadron states. We also refer to Refs.~\cite{Weinberg:2013cfa,Lucha:2019pmp}, which generally investigate how to apply the method of QCD sum rules and the large $N_c$ approximation to study multiquark states. One can use the Fierz transformation to write a local ``molecular'' current $[\epsilon^{abc} u_a d_b c_c] [\bar c_d u_d]$ as a combination of local diquark-diquark-antiquark currents $\epsilon^{fge} [\epsilon^{abf} u_a d_b] [\epsilon^{cdg} c_c u_d] \bar c_e$, and vice versa. However, this is an overall connection, {\it i.e.}, a ``molecular'' current can be written as a combination of many diquark-diquark-antiquark currents. We recommend interested readers to Ref.~\cite{Chen:2006hy}, where we first pointed out such connection by systematically studying the relation among various tetraquark currents. Take $\psi_{2\mu}$ defined in Eq.~(\ref{def:psi2mu}) as an example, we can transform it to be
\begin{eqnarray}
\psi_{2\mu} &=& [\epsilon^{abc} (u^T_a C \gamma_\mu u_b) c_c] [\delta^{de} \bar c_e \gamma_5 d_d]
\\ \nonumber &=& {1\over8} ~ \epsilon^{fge} [\epsilon^{abg} u^T_a C \gamma_\mu u_b] [\epsilon^{cdf} d_c^T C c_d] [\gamma_5 C \bar c_e^T]
\\ \nonumber &+& {1\over8} ~ \epsilon^{fge} [\epsilon^{abg} u^T_a C \gamma_\mu u_b] [\epsilon^{cdf} d_c^T C \gamma_5 c_d] [C \bar c_e^T]
\\ \nonumber &-& {1\over8} ~ \epsilon^{fge} [\epsilon^{abg} u^T_a C \gamma_\mu u_b] [\epsilon^{cdf} d_c^T C \gamma_\nu c_d] [\gamma^\nu \gamma_5 C \bar c_e^T]
\\ \nonumber &+& {1\over8} ~ \epsilon^{fge} [\epsilon^{abg} u^T_a C \gamma_\mu u_b] [\epsilon^{cdf} d_c^T C \gamma_\nu \gamma_5 c_d] [\gamma^\nu C \bar c_e^T]
\\ \nonumber &+& {1\over16}~ \epsilon^{fge} [\epsilon^{abg} u^T_a C \gamma_\mu u_b] [\epsilon^{cdf} d_c^T C \sigma_{\nu\nu^\prime} c_d] [\sigma^{\nu\nu^\prime} \gamma_5 C \bar c_e^T]
\\ \nonumber &+& {1\over32}~ \epsilon^{fge} [\epsilon^{acg} u^T_a C \gamma_\mu c_c] [\epsilon^{dbf} d_d^T C u_b] [\gamma_5 C \bar c_e^T]
+ \cdots
\\ \nonumber &+& {1\over32}~ \epsilon^{fge} [\epsilon^{adg} u^T_a C \gamma_\mu d_d] [\epsilon^{cbf} c_c^T C u_b] [\gamma_5 C \bar c_e^T]
+ \cdots \, ,
\end{eqnarray}
where $\cdots$ are other $\epsilon^{fge} [\epsilon^{acg} u^T_a C \Gamma_i c_c] [\epsilon^{dbf} d_d^T C \Gamma_j u_b] [\Gamma_k C \bar c_e^T]$ and $\epsilon^{fge} [\epsilon^{adg} u^T_a C \Gamma_l d_d] [\epsilon^{cbf} c_c^T C \Gamma_m u_b] [\Gamma_n C \bar c_e^T]$ components (this equation needs to be further simplified, but we shall not do this in the present study). This complicated relation suggests that although $\psi_{2\mu}$ can still be interpreted as a diquark-diquark-antiquark current containing many diquark-diquark-antiquark components, it seems much more natural to simply describe it as a ``molecular'' current.
In this sense, we can extract some useful information from the ``molecular'' currents being used:
\begin{itemize}

\item The $P_c^+(4312)$ can be described by the current $\psi_2$. The quark contents inside $\psi_2$ can be naturally separated into two color-singlet components, $[\epsilon^{abc} (u^T_a C \gamma_\mu u_b) \gamma_\mu \gamma_5 c_c]$ and $[\bar c_d \gamma_5 d_d]$. They are the two standard charmed baryon and charmed meson interpolating fields, which couple to $\Sigma_c^{++}$ and $\bar D^-$, respectively. Accordingly, $\psi_2$ would couple to the bound state of $[\Sigma_c^{++} \bar D^-]$ with $J^P = 1/2^-$, if it exists.
    Note that we made a typo in Ref.~\cite{Chen:2016otp} to label this as $[\Sigma_c^* \bar D]$. Hence, our result suggests that the $P_c^+(4312)$ can be well interpreted as the $[\Sigma_c^{++} \bar D^-]$ bound state with $J^P = 1/2^-$.

\item The $P_c^+(4440)$ and $P_c^+(4457)$ can be described by the currents $\xi_{33\mu}$ and $\psi_{2\mu}$. The quark contents inside $\xi_{33\mu}$ can be separated into $[\epsilon^{abc} (u^T_a C \gamma_\nu d_b) \gamma_\nu \gamma_5 c_c]$ and $[\bar c_d \gamma_\mu u_d]$, coupling to $\Sigma_c^+$ and $\bar D^{*0}$, respectively; while those inside $\psi_{2\mu}$ can be separated into $[\epsilon^{abc} (u^T_a C \gamma_\mu u_b) c_c]$ and $[\bar c_d \gamma_5 d_d]$, coupling to $\Sigma_c^{*++}$ and $\bar D^{-}$, respectively \big(the standard interpolating current coupling to $\Sigma_c^{*++}$ is $[\epsilon^{abc} (u^T_a C \gamma_\nu u_b) (g^{\mu\nu} - \gamma^\mu\gamma^\nu/4 ) c_c]$~\cite{Chen:2008qv}\big). Hence, our result suggests that the $P_c^+(4440)$ and $P_c^+(4457)$ can be well interpreted as $[\Sigma_c^{+} \bar D^{*0}]$ and $[\Sigma_c^{*++} \bar D^{-}]$ bound states with $J^P = 3/2^-$.

\item The current $\xi_{14}$ can also be used to describe one of the $P_c^+(4440)$ and $P_c^+(4457)$. Its quark contents can be separated into $[\epsilon^{abc} (u^T_a C \gamma_\mu d_b) \gamma_\mu \gamma_5 c_c]$ and $[\bar c_d \gamma_5 u_d]$, coupling to $\Sigma_c^{+}$ and $\bar D^0$, respectively. Hence, one of the $P_c(4440)$ and $P_c(4457)$ may be interpreted as the $[\Sigma_c^{+} \bar D^0]$ bound state with $J^P = 1/2^-$. The current $\xi_{14}$ is similar to $\psi_2$, but their extracted sum rule results are much different, simply because the two $up$ quarks inside $\xi_{14}$ are located in both of the two color-singlet components, so that there can be $up$ quark exchange between these two components, {\it i.e.}, Feymann diagrams exchanging $up$ quarks.

\item The current $\xi_{14}$ can be used to roughly describe one of the $P_c^+(4440)$ and $P_c^+(4457)$. Its quark contents suggest that one of the $P_c(4440)$ and $P_c(4457)$ may be interpreted as the $[\Sigma_c^{*+} \bar D^{*0}]$ bound state with $J^P = 5/2^-$.

\item There is still a place for the $P_c(4380)$, that is to use $\psi_{9\mu}$, whose quark contents can be separated into
\\ $[\epsilon^{abc} (u^T_a C \gamma_\nu u_b) \gamma_\nu \gamma_5 c_c]$ and $[\bar c_d \gamma_\mu d_d]$, coupling to $\Sigma_c^{++}$ and $\bar D^{*-}$, respectively. Again, its extracted sum rule result is much different from $\xi_{33\mu}$, due to the locations of the two $up$ quarks.

\end{itemize}

$\\$
{\it Mass splitting between the $P_c(4440)$ and $P_c(4457)$.}
Among the above three interpretations, there is still a problem in the third interpretation, that the two mass values extracted from the two currents $\xi_{33\mu}$ and $\psi_{2\mu}$, both of $J^P = 3/2^-$, are very close to each other (4.45~GeV and 4.46~GeV), so they can couple to the same physical state, although they have different internal structures. To check whether $\xi_{33\mu}$ and $\psi_{2\mu}$ couple to the same state or not, we calculate their off-diagonal correlation function~\cite{Chen:2018kuu}:
%
\begin{eqnarray}
\nonumber \Pi^{\xi_{33\mu}\psi_{2\mu}}_{\mu\nu}(q^2) &\equiv& i \int d^4x e^{iqx} \langle 0 | T \xi_{33\mu}(x) { \psi_{2\nu}^\dagger } (0) | 0 \rangle
\\ \nonumber &=& \left(\frac{q_\mu q_\nu}{q^2}-g_{\mu\nu}\right) (q\!\!\!\slash + M^*) \Pi^{\xi_{33\mu}\psi_{2\mu}}\left(q^2\right)
\\ && ~~~~~~~~~~~~~~~~~~~~~~~~~~~~~~~~ + \cdots \, ,
\end{eqnarray}
%
where $\Pi^{\xi_{33\mu}\psi_{2\mu}}\left(q^2\right)$ is contributed by the spin $3/2$ components of $\xi_{33\mu}$ and $\psi_{2\mu}$, while contributions from their spin $1/2$ components are all contained in $\cdots$.

\begin{figure}[!hbt]
\begin{center}
\includegraphics[width=0.4\textwidth]{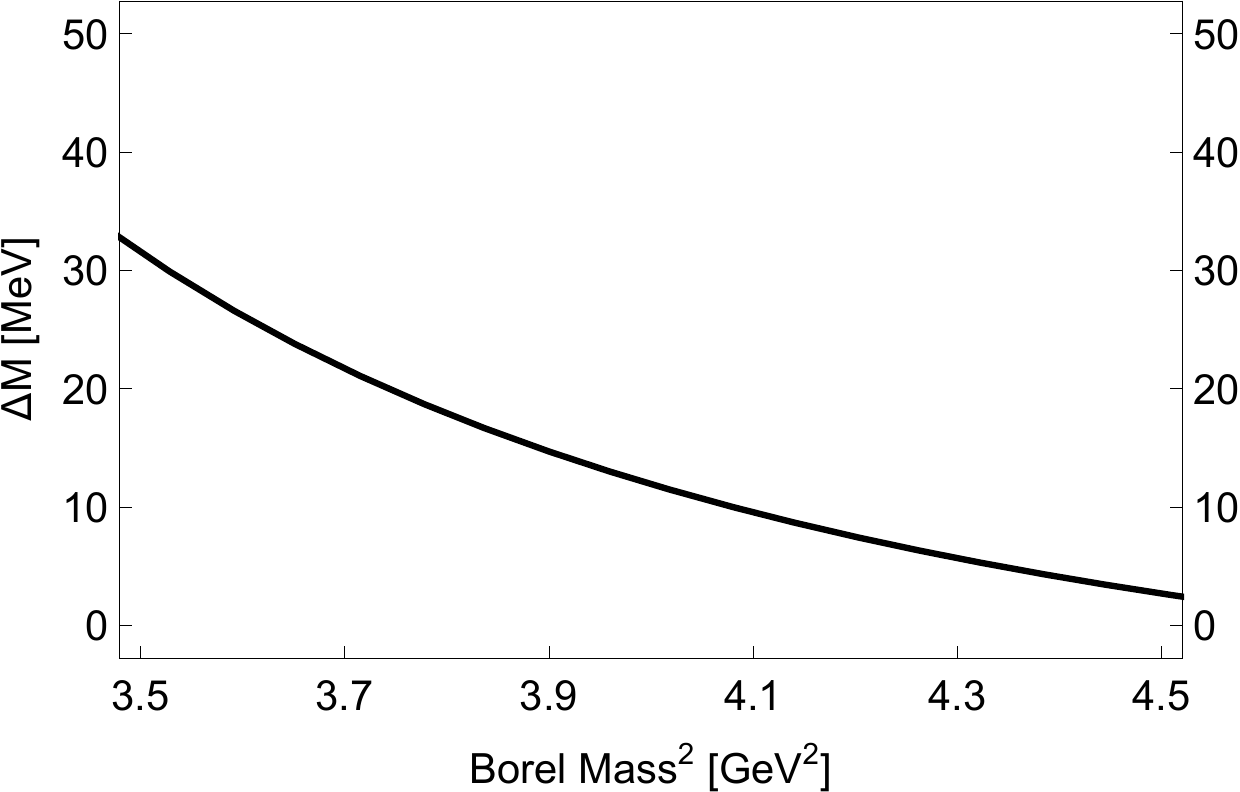}
\caption{The mass difference $\Delta M = M_{\xi_{33\mu}} - M_{\psi_{2\mu}}$ with respect to the Borel mass $M_B$. $M_{\xi_{33\mu}}$ and $M_{\psi_{2\mu}}$ are the masses extracted from the currents $\xi_{33\mu}$ and $\psi_{2\mu}$, respectively.}
\label{massplitting}
\end{center}
\end{figure}

If $\xi_{33\mu}$ and $\psi_{2\mu}$ do strongly couple to the same physical state $P_c^*$ with the mass $M^*$, we would have
\begin{eqnarray}
\langle 0 | \xi_{33\mu}(x) | P_c^* \rangle \langle P_c^* | { \psi_{2\nu}^\dagger } (0) | 0 \rangle \neq 0 \, ,
\end{eqnarray}
so that $\Pi^{\xi_{33\mu}\psi_{2\mu}}\left(q^2\right)$ should be nonzero. However, our QCD sum rule calculation gives us that
\begin{eqnarray}
\Pi^{\xi_{33\mu}\psi_{2\mu}}\left(q^2\right) = 0 \, .
\end{eqnarray}
Therefore, $\xi_{33\mu}$ and $\psi_{2\mu}$ should couple to different states, and can be used to describe the $P_c(4440)$ and $P_c(4457)$ at the same time. Since the two QCD sum rule parameters, the threshold value $s_0$ and the Borel mass $M_B$, are almost the same when investigating these two currents, we can extract their mass difference to be
\begin{eqnarray}
\Delta M = M_{\xi_{33\mu}} - M_{\psi_{2\mu}} = 8.1 ^{+30.9}_{-18.9} {~\rm MeV}\, ,
\end{eqnarray}
where the central value corresponding to $s_0 = 22$~GeV$^2$ and $M_B = 4.17$~GeV$^2$. The uncertainty comes from the Borel mass $M_B$, the threshold value $s_0$, the charm quark mass, and various condensates~\cite{Chen:2016otp}. It is much smaller than those of the absolute mass values listed in Table~\ref{tab:result}, although still significant. For completeness, we show $\Delta M$ as a function of $M_B$ in Fig.~\ref{massplitting}.

Anyway, the above mass splitting is consistent with the mass difference between the $P_c(4440)$ and $P_c(4457)$, making the third interpretation slightly more natural than others. It also suggests that the $P_c^+(4440)$ is preferred to be interpreted as the $[\Sigma_c^{*++} \bar D^{-}]$ bound state with $J^P = 3/2^-$, while the $P_c^+(4457)$ as the $[\Sigma_c^{+} \bar D^{*0}]$ bound state with $J^P = 3/2^-$.

$\\$
{\it Conclusion.}
In summary, after the discovery of the $P_c(4380)$ and $P_c(4450)$ by LHCb in 2015~\cite{Aaij:2015tga}, the $P_c(4312)$, $P_c(4440)$, and $P_c(4457)$ observed in the new LHCb experiment~\cite{lhcb} brought us a great surprise once more.
The coincidence of their measured masses with our previous theoretical predictions~\cite{Chen:2016otp} drives us to the ``molecular'' picture that the $P_c(4312)$ is the $[\Sigma_c^{++} \bar D^-]$ bound state with $J^P = 1/2^-$, and the $P_c(4440)$ and $P_c(4457)$ are the $[\Sigma_c^{*++} \bar D^{-}]$ and $[\Sigma_c^{+} \bar D^{*0}]$ bound states with $J^P = 3/2^-$, respectively. In this letter we further calculate the off-diagonal correlation function between $\xi_{33\mu}$ and $\psi_{2\mu}$, and their non-correlation confirms that we can extract two mass predictions from them for two states both having $J^P = 3/2^-$, whose mass difference is extracted to be $8.1 ^{+30.9}_{-18.9}$~MeV.

Besides the above picture, one of the $P_c(4440)$ and $P_c(4457)$ can also be interpreted as the $[\Sigma_c^{+} \bar D^0]$ bound state with $J^P = 1/2^-$ or the $[\Sigma_c^{*+} \bar D^{*0}]$ bound state with $J^P = 5/2^-$. There is still a place for the $P_c(4380)$, that is to be interpreted as the $[\Sigma_c^{++} \bar D^{*-}]$ bound state with $J^P = 3/2^-$. There exist more possible interpretations with the positive-parity assignments~\cite{Chen:2015moa,Xiang:2017byz}. To clearly understand their nature, one still needs further experimental information.

In the present QCD sum rule studies we intend to use various internal structures of hidden-charm pentaquark states to explain the $P_c(4312)$, $P_c(4440)$, and $P_c(4457)$ at the same time, while there are many other possible approaches. For example, in the molecular picture within the one-boson-exchange model, a beautiful picture is to interpret them as loosely bound $\Sigma_c \bar D$ molecular state with $J^P = 1/2^-$, $\Sigma_c \bar D^*$ with $J^P = 1/2^-$ and $\Sigma_c \bar D^*$ with $J^P = 3/2^-$, respective~\cite{Wu:2012md,liu}. However, the $[\Sigma_c \bar D^*]$ bound state with $J^P = 1/2^-$ was not investigated in the present study, and we shall study this possibility in the near future. At the same time, we shall study the $[\Sigma_c^* \bar D]$ bound state with $J^P = 1/2^-$ and the $[\Sigma_c^* \bar D^*]$ bound states with $J^P = 1/2^-$ and $3/2^-$, which were not investigated in the present study neither. We believe that a systematical QCD sum rule study might help to better understand these structures as well as this method itself.


We propose to measure the spin-parity quantum numbers of hidden-charm pentaquark states to verify whether the picture of the present study is correct or not. If it is correct, one would think that the internal structure of hadrons does influence their observed properties, and we might face the same situation as the light spectrum described by QED~\cite{Chen:2016spr}, so that lots of new exotic structures could be waiting to be discovered in the future. To end this letter, we would like to note that, together with many charmonium-like $XYZ$ states~\cite{pdg}, the hidden-charm pentaquarks are opening a new window for studying exotic hadronic matter and improving our understanding of QCD.

\section*{Acknowledgments}

This project is supported by
the National Natural Science Foundation of China under Grants No. 11575008, No. 11722540, No. 11621131001,
the 973 program,
the Fundamental Research Funds for the Central Universities,
and
the Chinese National Youth Thousand Talents Program.

\end{document}